# 19. Iteration and Co-design of a Physical Web Application for Outdoor Activities with Older Adults

Fatima Badmos[1], Emma Murphy[1], Michael Ward[1,] Damon Berry[1]
[1]Technological University Dublin, Dublin, Ireland.
Corresponding. Author@institution.ie

## Abstract

Existing research and physical activity guidelines highlight the benefits of outdoor physical activities for ageing populations. There is potential for technology to facilitate outdoor activity through Physical Web infrastructure. We proposed that embedding Physical Web applications that are engaging and interactive in public open spaces as part of interactive wellness parks can encourage older adults to participate in physical activities outdoors and motivate rehabilitation. We have created an initial design prototype based on design requirements generated from a qualitative field study with 24 older adults to explore their perceptions, experiences, and routines of outdoor physical activities. In this paper, we present an initial prototype and findings from a co-design session with 12 older adults, eliciting their feedback on the design and their ideas for future design iterations.

## Introduction

The world population is ageing at an increasing rate, and according to the World Health Organisation (WHO), the number of people aged 60 years and above outnumber children younger than five years [1]. As we age, we are more likely to experience age-related health conditions. Common age-related conditions among older adults include dementia, depression, osteoarthritis, hearing loss, and cognitive decline [7]. Research has shown that regular physical activities among older adults profoundly affect their well-being and can positively contribute to their mental and physical health [2]. This age group's lack of physical activities can result in a sedentary lifestyle [8]. Beyond the physical benefits, physical activities are essential in reducing the risk of cognitive decline associated with ageing and maintaining cognitive function. Additionally, physical activities improve sleep quality [9] and boost overall health and quality of life for older adults [3] Therefore, it is essential to develop community programmes that encourage older adults to engage in regular physical activities because, without adequate strategic planning, the healthcare burden associated with the increase in the number of older adults globally will significantly challenge the healthcare system. [4]

Several global initiatives for active ageing promote physical activities among older adults. According to the World Health Organization (WHO), active ageing is "the process of promoting health, social security, and social contribution of the elderly to promote their quality of life" [5]. Supporting older adults in regular activities is a crucial strategy to foster healthy and active ageing, a general strategy for maintaining physical and spiritual health [6]. There is wide support in the literature for the health benefits for older adults who engage in outdoor physical activity [10;12;11]. Benefits include increased participation in physical activity and the potential for increased social interactions. Creating outdoor mobility opportunities for older adults can improve their quality of life through increased opportunities for physical activity, promoting independence [13]. Therefore, the availability of outdoor spaces within communities of older adults is vital in promoting active ageing, physical activities, and social interaction.

## Need for co-design.

Various applications are designed for outdoor physical activities as mobile technologies evolve. These applications include walking apps [14], wearables [15], and exergaming [16]. Research has found that these interventions significantly increased physical activity and reduced sedentary behaviour [17].

However, implementation of these technological interventions has encountered resistance and underuse by older adults [18]. The most common problems include a lack of accessibility and usability for older adults with limited digital literacy [30]. The explanation for these issues could be a lack of consideration for older adults' needs and preferences, and most of these technologies are commercially made for older adults but were not designed with them, nor were older adults involved in the technology development process. Campelo and Katz [19] and [20; 22] reported a positive perception of using technology among older adults despite a lack of familiarity with technology for physical activities. However, Ivan et al. [31] analyse the mutual relationship between ageist stereotypes and technology and conclude that technology, developed by young people with the youth market in mind, produces prototypes that are more difficult for older people to use. Therefore, there is a need to investigate more tailored and accessible applications for outdoor physical activities involving older adults in the design process through co-design. Co-design is a methodology where the user participates in the design process of an application or service as an active co-designer [23]. A codesign-based approach has great potential for the proposed physical web interface. By understanding older adults' needs, preferences, and lived experiences and applying them within an iterative design method, we can co-create an interface that meets their needs. Additionally, this can help with creating an engaging and inclusive design.

### Digital Skills Divide Among Older Adults

As older adults become increasingly reliant on digital technologies that young people usually design without their participation, the ability to navigate new technologies has become essential for full participation in contemporary life. However, a persistent digital divide exists. Research has shown that older adults are not a homogenous group and should not be treated as such when designing technology for their needs [24]. While some older adults effortlessly integrate technology into their daily routines, others struggle to adapt, hindered by a lack of experience, interest, health, or access to resources [25]. Recognising the complexities within the older adult population is crucial in addressing these digital differences. There are many challenges in understanding exactly why the age-related digital skills gap appears among this demographic. According to Beneito-Montagut [26], individuals' prior technological experiences significantly influence their ease of navigating new technology. Research has shown that older adults who actively use technology during their professional lives often exhibit higher proficiency, having developed extensive digital skills. Many others in this age group proactively keep up with technological advancements out of personal interest or necessity. However, even experienced users may encounter challenges due to the rapid pace of technological change. Keeping abreast of the latest trends and adapting to new interfaces can be daunting, especially post-retirement, when access to technology may diminish [29]. In contrast, older adults with no prior exposure to technology face even greater barriers to accessing digital information. Motivation to acquire digital skills may be lacking, compounded by limited access to resources and distrust of online platforms [28]. It is crucial to recognise that most older adults, regardless of their past experiences, encounter difficulties in learning and engaging with digital technologies [29]. However, age-related physical, sensory, and cognitive impairments can further intensify these challenges and exacerbate the digital divide [27]. Age-related declines in vision, hearing, motor skills, and cognitive abilities can make it challenging for older adults to use digital technologies effectively.

### Design Requirements

At the beginning of this research, focus group workshops were organised to understand the perspectives of older adults. This included their experiences and perceptions of outdoor physical activities and the kind of activities they will be willing to engage in outdoors. Based on the findings from workshops with twenty-four older adults [21]. The following users' preferences were identified for the initial prototype design: creating an outdoor application that older adults with minimal technology skill or interest can engage with (no accompanying app needed to be downloaded), group activities to support socialisation, dance, music activities, and walking. Two essential parts were considered for the development of the prototype. The first was user engagement, designing the outdoor activities for users to engage in, tailored to group or individual physical activities. Second, the technology and the hardware to develop that technology. Physical Web technology was used to create user interaction. The prototype consisted of a user interface with two options: users listen and dance to their favourite music or tour different historic trees around the Technological University Dublin, Grangegorman Campus, which was originally built in the 18th century. An embedded Physical Web user interface was created using components such as the Arduino Nano IoT, the DFPlayer and MQTT. A key characteristic of the Physical Web is that it allows users to access web resources such as multimedia files without requiring them to install a mobile app [32].

Initial Design of Prototype Based on Design Requirements

**Dancing and Music Activities:** The first activity is the user interaction with the smart post embedded with a QR Code as shown in picture 1-3. On scanning the QR Code, the user is presented with a webpage with the option of choosing the activities they are interested in. If the users pick music, they are asked to select the music they are interested in from the list of music. Subsequently, they can enjoy listening to music directly from the smart post, eliminating the need for their personal phones or downloading additional applications. Additionally, users have the freedom to dance to the music without relying on their mobile devices. The novelty of the interaction stems from the fact that the participants are listening from the artefacts in the physical environment rather than from their phones. However, the prompt to play the music is enacted from their mobile devices.

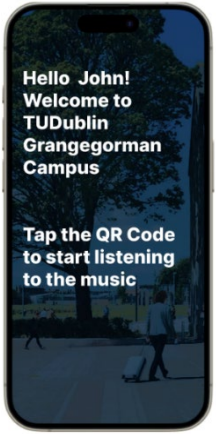

*Figure 19-1: Web interface*

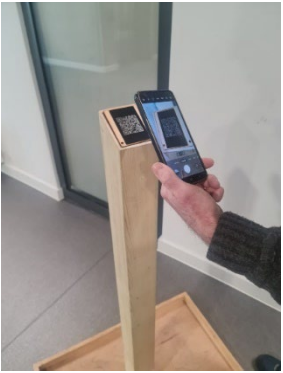

*Figure 19-2: User scanning the QR code music.*

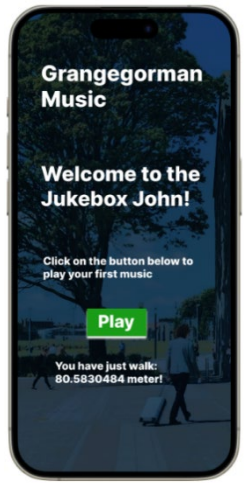

*Figure 19-3: Web interface to initiate the music*

**History Tours:** Following the previous activity is the history tour experience. Upon selecting the history tour option from the web page, users can simply tap the QR code located on the artefact in the physical environment. This action presents them with two choices: they can either listen to the history of the place, which include details about the year it was built, story surrounding the place through the technology integrated around the building, or they can read the information via the web on their mobile phones.

## Exploration and Co-design Session

A co-design workshop was organised with 12 older adults at the Technological University Grangegorman Campus in Dublin. This co-design session aimed to:

1. Explore perceptions and experiences of the group's outdoor activities.
2. Elicit feedback on the initial prototype presented above.
3. use this initial prototype to ideate and iterate on the application design.

## Method

### Participants

Participants were recruited from a voluntary service organisation, which provides a range of social activities to the local older adult community in Dublin. Inclusion criteria for participating in the session were that they were over 60 years old, could give informed consent, and were willing to try out new technology and discuss their experience of engaging in outdoor activities. Twelve older adults took part in the workshop, with ages ranging from 60-100. Table 1.

Table 19-1: Participants' Demographic Table

| P. No | Gender | Age Range | How often do you engage in Physical activity outdoors a week? | Do you currently have a working mobile phone? | Does your phone have internet? | What do you use your phone for mostly? | Have you ever used your phone to scan a QR code | Can you use the on-screen Keyboard on your phone to type? |
|---|---|---|---|---|---|---|---|---|
| 1 | Male | 81-90 | Everyday | No | No | Receive and make calls | No, | No |
| 2 | Male | 81-90 | Three Times | Yes | yes | Receive and make calls, send messages, social media, Google map | No | No |
| 3 | Male | 81-90 | Three Times | No | No | No Mobile phone | No | No |
| 4 | Female | 60-70 | Everyday | Yes | Yes | Receive and make calls, send messages, social media, Google map | Yes | Yes |
| 5 | Male | 81-90 | Three Times | Yes | No | Receive and make calls | No | No |
| 6 | Female | 71-80 | Everyday | Yes | No | Receive and make calls, send messages | No | Yes, slowly |
| 7 | Female | 90-100 | Everyday | Yes | No | Receive and make call | Yes | No |
| 8 | Female | 85 | Everyday | Yes | No | Receive and make call | No | Yes |
| 9 | Female | 81- 90 | Everyday | Yes | Yes | Receive and make calls, send text messages, social media | Yes | Yes |
| 10 |  | 81-90 | Female | Everyday | Yes | Receive and make calls, send messages | No | No |
| 11 | Female | 89 | Seldom | Yes | Yes | Receive and make calls, send text messages, social media, Google map | No | Yes, slowly |
| 12 | Female | 81-90 | Everyday | Yes | No | Receive and make calls, send text messages | No | Not |

Participants travelled from their activities centre to the university for the workshop. They were divided into three focus groups. Demographic questionnaires were given to the participants to capture their age, how often they engage in outdoor activities, what kind of mobile phone they have, and what they mainly use their mobile phones for. The workshop protocol was developed to explore the initial findings with the participants, introduce the early prototype, and get feedback for further prototype iterations. The workshop lasted approximately two hours. The workshop was divided into two sections; the first was to explore the group's perceptions and experiences about engaging in outdoor activities. A semi-structured interview protocol included some questions from previous workshops under the following themes:

- Types of outdoor physical activities the participants currently engaged in.
- What activities would they prefer to do outdoors?
- Motivation, facilitators, and barriers to engaging in outdoor physical activities.

In the second stage of the workshop, participants were introduced to the Physical Web with explanations and a video showing how technology like QR codes can be used. The facilitators observed the participants' interactions to identify usability and accessibility issues experienced during the process. After the participants interacted with the prototype, feedback and ideas for the next iteration were collected. Participants' inputs were audio recorded, and the group facilitator also used Post-it papers to record participants' reflections.

## Analysis

The focus group interactions were audio-recorded and transcribed verbatim. Two researchers reviewed each transcript to ensure the recorded and anonymised identifiable data was accurate according to Braun and Clarke's [33] inductive data analysis framework; post-its data were arranged according to the themes that emerged from the data.

## Findings

Four major themes were identified in the data analysis: Ease of prototype use, First impression of the prototype, prototype improvement suggestion, Privacy, and security concern. A summary of these major themes and their subthemes is presented in Table 2.

| Major Themes | Sub Themes | Quotes |
|---|---|---|
| Ease of Use | Limited Understanding of QR Codes functions | "Hard to understand", "Difficult to get how it works", "Beyond comprehension for older users", "Cannot be used without smart phone", "Where is the music coming from?" |
| First Impressions and Experience of the prototype | Impressive<br><br>Perceived complexity and Age-Related barriers | "Love the experience", "Love the music features", "Love the interaction for outdoor activities", "It will help with social engagement", QR code outside will be good to get easy access to information, "We are too old to be bothered with new technology", "This is for young people", "Too complex for older people to understand |
| Prototype Improvement suggestions | Simpler interactions | "Make it easier to use", "For the music application, enable automatic playback minimal interaction", "Consider alternative forms of interaction like loyalty cards or tags", "Simplify with one-button operation", "Consider a smaller device or phone integration", "Enable automatic playback with minimal interaction", "Key holder idea or button for easier interaction" |
| Privacy and Security Concerns | Stolen data | "Who has access to our data?" "How is the technology made?", How secure is that thing? |

**Ease of Use:** A significant finding was the varied levels of understanding among participants regarding how QR codes functioned. Several participants who did not have smartphones found the process particularly challenging and expressed confusion and uncertainty about the mechanics of scanning QR codes. This confusion may stem from limited exposure to QR code scanning or difficulties in understanding the instructions provided, indicating a need for more intuitive and inclusive interfaces to enhance user comprehension and interactions. Another important aspect of ease of use is related to the novelty of hearing audio from the physical environment rather than through participants' mobile phones when they click play, highlighting a lack of familiarity with this technology. For many older adults in this cohort, this experience was unfamiliar and unexpected, leading to confusion and uncertainty about how to interpret the audio feedback. Some participants expressed surprise at hearing sounds emanating from the physical environment, highlighting a shift in their accustomed modes of audio interaction. This novelty factor may have contributed to initial hesitation or difficulty in engaging with the physical web application, particularly for those who were not expecting this mode of audio playback.

**Impression of the Prototype:** Despite the mixed reactions to the mechanics of how the prototype works, several participants reported enjoying the experience of interacting with the prototype. They appreciated the opportunity to engage in outdoor activities while leveraging digital technology to enhance their physical and social experience. For these participants, the technology provided an immersive way to interact with their surroundings, enriching their outdoor experiences and fostering a sense of enjoyment and fulfilment. However, while some of the participants described the interaction with the prototype as engaging, they perceived the technology as suitable for younger audiences as they could not be bothered to learn how to use it.

**Prototype Improvement suggestions:** The findings from the focus group workshop emphasise the importance of enhancing the usability and accessibility of the physical web application for older adults. Seven Participants highlight the necessity to "make it easier to use" through various suggestions and recommendations. Firstly, for the music application, participants suggested enabling automatic playback with minimal user interactions and inputs. This feature would allow users to enjoy audio content without the need for manual input, enhancing the user experience and reducing cognitive overload. Additionally, participants suggested considering alternative forms of interaction, such as loyalty cards or tags. These alternative methods would provide older adults with options beyond traditional smartphone interfaces, catering to diverse preferences and mobile and technological literacy levels. Another suggestion involved considering a smaller device or phone integration. Participants highlighted the importance of portability and seamless integration with existing devices, facilitating adoption and usage among older adults. Moreover, enabling automatic playback with minimal interaction was reiterated as a key enhancement. By reducing the need for manual input, older adults can engage with the physical web applications effortlessly, enhancing their overall user experience. Finally, the idea of incorporating a key holder or button for easier interaction was proposed. This physical addition would provide older adults with a tangible interface for accessing the features of the Physical Web application, further enhancing usability and accessibility. Furthermore, simplifying the interface with a one-button operation emerged as a prominent suggestion. By simplifying the interaction process, older adults would be able to navigate the Physical Web application more intuitively, promoting user interaction and satisfaction.

**Privacy and Security Concerns:** The fear of security issues emerged as a common theme among participants. Older adults expressed reservations about the safety of using QR codes, indicating a need for robust security measures and clear communication to address these concerns and build trust in the technology. Three of the participants raised questions such as "Who has access to our data?", "where is the data stored?" indicating a lack of understanding of the proposed system. Another aspect of concern was connected to the transparency of the technology's development process. One participant questioned, "How is the technology made?" expressing a desire for greater transparency and understanding of the underlying processes involved in creating the Physical Web application. Lastly, participants highlighted their doubts about the overall security of the technology, asking, "How secure is that thing?" This question reflects a vital concern about the integrity of the technology and its ability to safeguard users' information and ensure a secure user experience.

# Discussion

The diverse levels of understanding among participants regarding QR code functionality highlights a crucial area for improvement in the design of the Physical Web application. Clearer instructions and more intuitive interfaces are necessary to address confusion and uncertainty, ensuring that older adults can easily navigate the technology [34]. By prioritising user comprehension, interaction and inclusivity, designers can enhance usability and promote greater adoption of technology among this older adult cohort. Participants' belief that QR code technology is too complicated for older adults highlights the need for user-friendly design elements that cater to the diverse cognitive and technological literacy levels of this cohort of older adults [36]. Addressing age-related barriers requires simplifying interfaces, streamlining interaction processes, and providing alternative forms of interaction when needed. By making technology more accessible, inclusive, and intuitive, designers can overcome perceived complexities and promote inclusivity among older adults [35].

Participants' concerns about data access, technology transparency, and security underline the importance of prioritising privacy and security measures in the design and implementation of applications. Addressing these concerns requires transparent and clear communication with users on where their data is stored and the strategies in place to protect their data. Also, it is important to limit the amount of data collected for such interactions. By instilling confidence and trust, designers can alleviate apprehensions and promote greater acceptance of the technology among older adults. Participants' difficulties in using the technology highlight the need for streamlined interfaces and intuitive interaction processes. Features such as one-button interaction and automatic playback with minimal interaction can enhance usability and accessibility for older adults in this cohort. Additionally, considering alternative forms of interaction, such as loyalty cards or tags, can provide users of Physical Web, especially those without smartphones, with options beyond traditional smartphone interfaces. By prioritising ease of use, designers can create a more inclusive and user-friendly technology that meets the diverse needs and preferences of older adult users.

## Limitation

A limitation of this study is the small number of samples of older adults involved and their age range of 65-100, with most of the participants aged over 80. From a sample of 12 participants, only 3 participants reported using QR codes before the study, and most of the participants did not have a smartphone to scan a QR code. This sample may not fully represent the diverse population of older adults. The sample of the study limits the generalisability of the findings to broader populations of older adults with varying age groups, backgrounds, and digital experiences. Moreover, the study may not have fully reflected the specific contexts in which older adults would use the Physical Web application, which is in outdoor environments with varying levels of ambient noise or lighting conditions. This study was conducted in a study room with the older adults. Understanding the contextual factors like varying levels of ambient noise or lighting conditions that impact usability and engagement could inform more tailored design decisions. Lastly, our future iterative workshops will be organised outdoors to give the participants an environment to use the prototype. We will also include many participants with a diverse age range and digital skills and experience.

## Conclusion.

The research aimed to codesign a Physical Web application with older adults to encourage outdoor physical activity. The study presented in this paper includes a focus group workshop exploring the initial prototype of the application with 12 older adults. Participants without smartphones face challenges using the application, and other participants find the QR code functionality difficult to understand, while others question the security of the QR code technology. Despite these challenges, most participants found the technology useful for outdoor activities. Future iterations of the design will focus on considering the feedback from the participants, redesigning the interface for clarity, and exploring different modes of interaction for those without smartphones. However, there is a need to ensure that the Physical Web technology remains usable, inclusive and accessible for all older adults.

## Acknowledgements

This work was conducted with the financial support of the Science Foundation Ireland Centre for Research Training in Digitally Enhanced Reality (d-real) under Grant No. 18/CRT/6224. For the purpose of Open Access, the author has applied a CC BY public copyright licence to any Author Accepted Manuscript version arising from this submission.